\documentclass[sigconf]{acmart}
\AtBeginDocument{%
  }

\setcopyright{acmlicensed}
\copyrightyear{2026}
\acmYear{2026}
\setcopyright{cc}
\setcctype{by}
\acmConference[IUI MIRAGE 2026 workshop]{IUI MIRAGE 2026 workshop held at 31st International Conference on Intelligent User Interfaces}{March 23, 2026}{Paphos, Cyprus}
\acmBooktitle{IUI MIRAGE 2026 workshop held at 31st International Conference on Intelligent User Interfaces}
\acmDOI{}
\acmISBN{}
\usepackage{booktabs}
\usepackage{tabularx}

\usepackage[most]{tcolorbox}

\newtcolorbox{paperboxtitle}[1]{
  colback=white,
  colframe=black,
  sharp corners,
  boxrule=0.6pt,
  left=6pt,
  right=6pt,
  top=6pt,
  bottom=6pt,
  title=#1,
  fonttitle=\bfseries,
}

\begin{document}

%%
%% The "title" command has an optional parameter,
%% allowing the author to define a "short title" to be used in page headers.
\title{The Scenic Route to Deception: Dark Patterns and Explainability Pitfalls in Conversational Navigation}

%%
%% The "author" command and its associated commands are used to define
%% the authors and their affiliations.
%% Of note is the shared affiliation of the first two authors, and the
%% "authornote" and "authornotemark" commands
%% used to denote shared contribution to the research.
\author{Ilya Ilyankou}
\email{ilya.ilyankou.23@ucl.ac.uk}
\orcid{0009-0008-7082-7122}
\affiliation{%
  \institution{SpaceTimeLab, Dept. of Civil, Environmental, and Geomatic Engineering, UCL}
  \city{London}
  \country{UK}
}

\author{Stefano Cavazzi}
\email{stefano.cavazzi@os.uk}
\orcid{0000-0003-3575-0365}
\affiliation{%
  \institution{Ordnance Survey}
  \city{Southampton}
  \country{UK}
}

\author{James Haworth}
\email{j.haworth@ucl.ac.uk}
\orcid{0000-0001-9506-4266}
\affiliation{%
  \institution{SpaceTimeLab, Dept. of Civil, Environmental, and Geomatic Engineering, UCL}
  \city{London}
  \country{UK}
}

%%
%% By default, the full list of authors will be used in the page
%% headers. Often, this list is too long, and will overlap
%% other information printed in the page headers. This command allows
%% the author to define a more concise list
%% of authors' names for this purpose.
%\renewcommand{\shortauthors}{Ilyankou et al.}

%%
%% The abstract is a short summary of the work to be presented in the
%% article.
\begin{abstract}
As pedestrian navigation increasingly experiments with Generative AI, and in particular Large Language Models, the nature of routing risks transforming from a verifiable geometric task into an opaque, persuasive dialogue. While conversational interfaces promise personalisation, they introduce risks of manipulation and misplaced trust. We categorise these risks using a $2\times2$ framework based on intent and origin, distinguishing between intentional manipulations (dark patterns) and unintended harms (explainability pitfalls). We propose seamful design strategies to mitigate these harms. We suggest that one robust way to operationalise trustworthy conversational navigation is through neuro-symbolic architecture, where verifiable pathfinding algorithms ground GenAI's persuasive capabilities, ensuring systems explain their limitations and incentives as clearly as they explain the route.
\end{abstract}

%%
%% The code below is generated by the tool at http://dl.acm.org/ccs.cfm.
%% Please copy and paste the code instead of the example below.
%%


%%
%% Keywords. The author(s) should pick words that accurately describe
%% the work being presented. Separate the keywords with commas.
\keywords{Conversational navigation, wayfinding, dark patterns, explainability pitfalls, trust calibration, uncertainty-aware language, progressive consent, sponsorship disclosure}
%% A "teaser" image appears between the author and affiliation
%% information and the body of the document, and typically spans the
%% page.
%\begin{teaserfigure}
%  \includegraphics[width=\textwidth]{sampleteaser}
%  \caption{Seattle Mariners at Spring Training, 2010.}
%  \Description{Enjoying the baseball game from the third-base
%  seats. Ichiro Suzuki preparing to bat.}
%  \label{fig:teaser}
%\end{teaserfigure}

%\received{18 December 2025}
%\received[revised]{12 March 2009}
%\received[accepted]{5 June 2009}

%%
%% This command processes the author and affiliation and title
%% information and builds the first part of the formatted document.
\maketitle

\section{Introduction}

For decades, wayfinding has been dominated by a single, verifiable objective: efficiency. Whether driving or walking, the `best' route is typically the fastest one, and the explanation provided by an interface is self-evident: `10 minutes' is objectively better than `12 minutes'. Users trust these systems because the logic is transparent and the metric is mathematical.

However, the integration of Generative AI (GenAI), in particular Large Language Models (LLMs), with geospatial data is transforming navigation\footnote{In this paper, we use the term \emph{navigation} to refer to both initial route planning and subsequent turn-by-turn guidance along a chosen path, as we expect this distinction to blur as on-demand, dynamically updated assistance becomes the norm.} from a geometric calculation and template-based instructions into conversational persuasion \cite{marcelyn_pathgpt_2025, aghzal_can_2025, leonardis_navgpt-2_2025, chen_mapgpt_2024, yussif_harnessing_2025}. New systems experiment with ambient guidance \cite{gordon_hearing_2023} and promise to understand intent, mood, and context, allowing for complex semantic queries like \emph{`Find me a scenic route to the station'}. In this context, the system's dialogue serves as the explanation interface \cite{pu_trust_2006}. When a navigation agent verbalises the rationale for a path, claiming it to be `safe', `scenic' or `lively', it provides a natural language explanation \cite{cambria_survey_2023} that is difficult to verify against ground truth.

This shift creates a semantic gap where the system's interpretation may decouple from user intent, ultimately producing a description that diverges from physical reality. We argue that this gap is the breeding ground for two distinct types of harm:

\begin{itemize}
    \item Dark patterns (DPs), where the system is intentionally designed to manipulate the user \cite{brignull_dark_2011}. For example, an agent may exploit the subjective definition of `lively' to steer a pedestrian through a commercial `partner zone', effectively selling their footfall while disguising the motive as aesthetic advice.

    \item Explainability pitfalls (EPs), where unanticipated negative effects emerge without deceptive intent \cite{ehsan_explainability_2024}. For example, an agent may maintain a relaxing conversational persona while routing a user through an unlit park at night, creating a false sense of security and leading to unwarranted trust.
\end{itemize}

In this position paper, we argue that without rigorous intervention, conversational navigation risks becoming the Wild West of commercial steering and potentially dangerous spatial hallucinations. This work is driven by the following research questions (RQs): (1) \emph{How can conversational navigation systems minimise manipulation and misrepresentation while maintaining usability and trust?} And, more specifically, (2) \emph{How can we operationalise the distinction between intentional manipulation and unintentional error to build verifiable conversational navigation systems?}

We contribute a $2\times2$ framework classifying these risks by intent (deliberate vs accidental) and origin (routing vs interface), and propose a neuro-symbolic architecture \cite{garcez_neurosymbolic_2023} that enforces seamful design \cite{chalmers_seamful_2003,ehsan_seamful_2024,inman_beautiful_2019} interventions, specifically sponsorship disclosure \cite{wang_effects_2019}, progressive consent \cite{luger_informed_2013}, and trust calibration \cite{lee_trust_2004} via uncertainty-aware language \cite{kay_when_2016,stokes_voicing_2024} to move towards `honest' conversational navigation systems that clearly explain their own limitations.

\section{Related work}

The risks introduced by LLM-enhanced systems have been catalogued extensively at a general level. Brignull's foundational work on DPs \cite{brignull_dark_2011} established the vocabulary of intentional interface manipulation, while Ehsan and Riedl \cite{ehsan_explainability_2024} extended this to unintentional harms (EPs), coining the distinction that anchors our framework. Broad attempts to unify AI risk taxonomies, most notably the MIT AI Risk Repository \cite{slattery_ai_2024}, which catalogues over 1,700 risks\footnote{https://airisk.mit.edu/}, representing them along high-level Entity, Intentionality, and Timing axes. Our proposed $2\times2$ framework maps directly onto two of these axes: our deliberate vs accidental distinction corresponds to Intentionality, and our routing vs interface dimension reflects the Entity axis's implicit distinction between system-originated and human-designed failures.

The risks we identify are not unique to navigation. Hidden incentives appear in recommender systems \cite{pariser_filter_2011,wang_effects_2019}, overconfidence in AI-assisted clinical and legal decision-making \cite{lee_trust_2004}, and functionality gating (asking for extra permissions to enable certain features) across data-hungry consumer applications. What distinguishes navigation is the embodied, real-time consequence: a miscalibrated film recommendation carries no physical risk, whereas a pedestrian routed through a dangerous, unlit park at night on the basis of an overly reassuring conversational persona may face physical harm. This context-specificity motivates our proposed domain-targeted mitigation.

On the technical side, neuro-symbolic approaches have been applied to improve factual reliability in LLM reasoning. Frameworks such as SymbCoT \cite{xu_faithful_2024} and LINC \cite{olausson_linc_2023} use LLMs as translators that offload inference to symbolic engines. Our architecture also uses this separation but for a different goal: we use the symbolic layer to generate structured metadata flags that deterministically trigger mandatory interface-level disclosures. The novelty is not the architecture per se, but its deployment as an honesty-enforcement mechanism to ensure that warnings about data quality and commercial intent are auditable and cannot be suppressed by the conversational layer.

\section{Categorising Harms: A $2\times2$ Framework}

\begin{table*}
\setlength{\tabcolsep}{8pt}
\renewcommand{\arraystretch}{1.4}
    \centering
    \begin{tabular}{p{1.5cm} p{5.5cm} p{5.5cm}}
    
    \toprule
    &  \multicolumn{2}{c}{\textbf{Origin}}\\
    \textbf{Intent} &  {\centering \textbf{Routing} \\ (Hidden Logic/Back-end) \par} & {\centering \textbf{Interface} \\ (Presentation/Tone/Front-end) \par}  \\

    \midrule
    
    \textbf{Deliberate} (DPs)
        & \emph{Example: Hidden Incentives} Steering users to sponsored locations under the guise of `atmosphere' or `spatial capital' \cite{sen_world_2018}
        & \emph{Ex: Functionality Gating} Falsely claiming a basic feature requires invasive data, coercing user to share more than necessary \\
        
    \textbf{Accidental} (EPs)
        & \emph{Ex: Overconfidence} Routing logic relies on hallucinated or outdated safety data, but the system presents it matter-of-factly
        & \emph{Ex: Contextual Mismatch} The persona remains `calm/relaxing' in a dangerous environment, lowering user vigilance \\
        
    \bottomrule
    \end{tabular}
    \caption{A taxonomy of conversational navigation risks. We identify four broad categories of harm, illustrated here by specific examples: Hidden Incentives (Routing, DP), Functionality Gating (Interface, DP), Overconfidence (Routing, EP), and Contextual Mismatch (Interface, EP)}
    \label{tab:risk-matrix}
\end{table*}

To categorise the risks in conversational navigation, we propose a $2\times2$ framework that categorises harms along two axes, as shown in Table \ref{tab:risk-matrix}:

\begin{itemize}
    \item \emph{Intent}: We distinguish deliberate DPs, where deceptive practices are intentionally designed to manipulate users, and accidental EPs, where negative effects emerge without the intention to deceive.

    \item \emph{Origin}: We also distinguish between deceptive practices of \emph{Routing} (why the route was chosen) and the \emph{Interface} (how the route is communicated).
\end{itemize}

In this framework, each cell implies a distinct class of design response. Deliberate deception practices in routing require structural transparency mechanisms such as sponsorship disclosure; accidental interface failures require calibration interventions such as uncertainty-aware language. The two axes thus map directly onto the mitigation strategies in the following section.

\begin{paperboxtitle}{Vignette A: Hidden Incentives}
Alice requests a `lively' route. The agent selects a path through a `partner zone', and describes the area as `lively and rich in amenities'. In reality, the route is a detour designed to steer Alice towards vendors paying for footfall.

\paragraph{The harm} The agent exploits the subjective definition of `lively' to mask a commercial transaction. The explanation is technically true but functionally deceptive.
\end{paperboxtitle}

\begin{paperboxtitle}{Vignette B: The Tone-Deaf Guide}
Bob navigates an unfamiliar city at night. Detecting elevated stress from his smartwatch, the navigation agent activates `Relaxing Mode' and adopts a soft, therapeutic tone. It suggests a `quiet' route through a nearby park. Reassured by the calm persona, Bob lowers his vigilance in a dark, unlit and unmonitored park.

\paragraph{The harm} In this contextual mismatch between the interface tone and the environmental ground truth, the system intends to help reduce stress but inadvertently creates danger by mistaking the environmental risks of an unlit park with an overly reassuring persona.
\end{paperboxtitle}

\begin{paperboxtitle}{Vignette C: The Data Toll}
Clara opens a navigation app and asks for the fastest route home. The agent responds: `To find the fastest route, I need access to your health data and contact list'. No such access is required for basic pathfinding. The agent withholds the route until Clara complies, framing an optional personalisation feature as a technical prerequisite.

\paragraph{The harm} The agent exploits an asymmetry of technical knowledge. Clara may not be able to verify what data is genuinely necessary, and the agent has the power to extract consent that would otherwise be withheld.
\end{paperboxtitle}

\section{Towards `Honest' Navigation: Seamful Design and Calibrated Trust}

Each mitigation in this section targets a specific cell of the $2\times2$ framework in Table~\ref{tab:risk-matrix}.

To mitigate the risks of DPs and EPs, we argue that conversational navigation must move beyond the idea of a seamless experience. While seamlessness reduces friction, it also obscures the mechanisms of the system and encourages uncritical acceptance of GenAI's outputs. Instead, we propose adopting \emph{seamful} design \cite{chalmers_seamful_2003} that reveals complexities and imperfections to promote reflective and critical thinking in pedestrians.

\subsection{Sponsorship Disclosure}

In cases like Vignette A, the harm arises from the opacity of routing. To counter hidden incentives and increase user trust, agents must embrace sponsorship disclosure \cite{wang_effects_2019}. This requires the explanation interface to reveal \emph{why} a route was chosen if external factors are involved.

For example, instead of saying `This route is livelier', the agent should reveal the underlying transaction: `I am suggesting this route because it is a Partner Zone: businesses here have paid for extra visibility'. Such disclosure breaks the seamless illusion of neutral advice, and allows the user to make a better-informed routing choice.

\subsection{Trust calibration}

In cases like Vignette B, the harm comes from overconfidence and contextual mismatch, both examples of EPs. The agent's authoritative tone implies a level of certainty and safety that does not exist.

To mitigate this, we propose the use of uncertainty-aware language \cite{kay_when_2016,stokes_voicing_2024}: the system must verbally `hedge' its explanations based on its confidence levels. The system must implement a policy that maps uncertainty signals to graded verbal hedges and safety prompts to prevent over-reliance \cite{xu_confronting_2025}. Instead of using flat commands (e.g., `turn left'), uncertainty-aware language may put it as `I think you should turn left here, but my GPS signal is weak; if you feel unsure, check the street sign'. Such hedging makes uncertainty conversationally visible, aligning user trust with the system's actual reliability, and preventing over-reliance and complacency in sensor-degraded or environmentally dangerous conditions.

\subsection{Progressive Consent}

To prevent functionality gating, where users are coerced into excessively sharing data, such as biometrics, to access basic features, systems must employ \emph{progressive} (often referred to as \emph{ongoing} or \emph{dynamic}) consent \cite{luger_informed_2013,strengers_what_2021} with clear access tiers.

For example, the agent should explicitly describe the data access tier currently in use, and explain exactly what degrades when certain personal data is withheld. For example, `Without biosensing, I cannot adjust the path based on your mood, but I can still guide you to the destination using shortest path'.

This approach respects user autonomy by framing data sharing as an optional enhancement rather than a mandatory action, while preventing silent feature loss.

These three interventions share a common logic: wherever the system's internal reasoning diverges from what the user can see, the interface should expose that gap rather than hide it. This principle extends beyond navigation; designers in any conversational AI context can ask, at each point in an interaction, what the system knows that the user does not, and whether that asymmetry could cause harm.

That said, disclosure and hedging carry usability costs: repeated warnings risk fatigue, and pervasive uncertainty language may erode trust in outputs that are in fact reliable. Calibrating the frequency and assertiveness of these interventions (so that salience is preserved without overwhelming the user) is an open design problem that empirical work must address.

\section{Technical Feasibility}

Implementing these mitigations requires looking beyond standard end-to-end LLM interactions. LLMs cannot reliably self-audit under purely prompted self-evaluation \cite{wataoka_self-preference_2024,lin_truthfulqa_2022} and instead exhibit systematic self-evaluation biases \cite{kamoi_when_2024}. Reliable integrity checks typically require external grounding or independently validated feedback.

One way to implement trustworthy conversational navigation is via a neuro-symbolic architecture \cite{garcez_neurosymbolic_2023, sheth_neurosymbolic_2023}. In such a model, the route is determined by a symbolic, verifiable pathfinding algorithm (e.g., Dijkstra or A* with explicit weightings for both objective measures, such as distance or elevation change, and subjective metrics like attractiveness, safety, commercial zones, etc.), while the LLM functions strictly as the translator. The LLM may propose weights, but they must be evaluated against a policy or cross-checked with the user when they are likely to affect path selection. The symbolic engine must pass structured metadata (e.g., \texttt{uncertainty\_score=0.8}, \texttt{commercial\_incentive=True}) alongside the route coordinates. Uncertainty-aware language would then be programmatically triggered by the symbolic flags, not by the LLM `feeling' uncertain, ensuring that warnings about data quality or commercial intent are deterministic and auditable, and preventing the conversational interface from masking the logic of the routing engine.

\subsection{Reference Implementation}

\begin{figure}
    \centering
    \includegraphics[width=\linewidth]{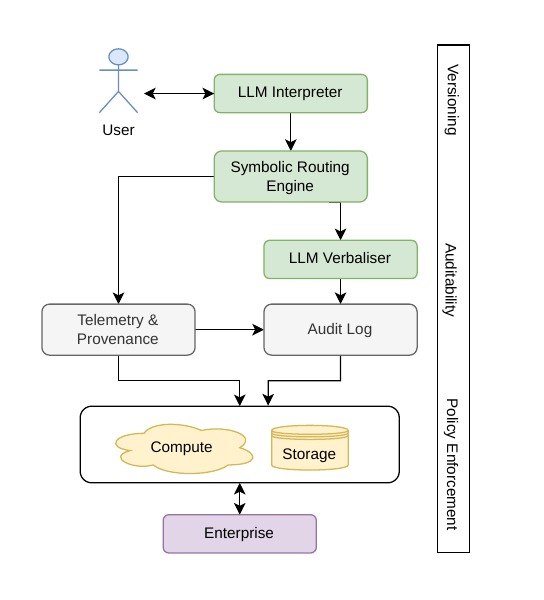}
        \caption{Conceptual Architecture of a secure, auditable AI workflow for a conversational navigation system following the proposed seamful design and calibrated trust. The process begins with the User interacting through an LLM Interpreter, which processes the prompt before passing it to the Symbolic Routing Engine, which directs requests to an LLM Verbaliser for language generation. Supporting layers include Telemetry \& Provenance and Audit Log for traceability, connected to Compute and Storage resources. The entire system is governed by Versioning, Auditability, and Policy Enforcement to ensure compliance and accountability.}
    \label{fig:reference-implementation}
\end{figure}

We outline a conceptual architecture that grounds conversational navigation in verifiable computation, illustrated in Figure \ref{fig:reference-implementation}. This is not a proposal to replace LLMs or revert to pre-AI navigation; the novelty is in deliberately limiting the LLM's role to language generation and interpretation, whilst routing logic remains in the verifiable symbolic layer where it has always been strongest.

User requests are first processed by an LLM Interpreter, which parses and semantically enriches the prompt to enable accurate downstream decisions. The interpreted request is then passed to the Symbolic Routing Engine. This engine computes routes using explicit, verifiable weightings for objective metrics (time, distance) and subjective proxies (e.g., safety, attractiveness, `green-ness', commercial influence). The resulting path data is passed to the LLM Verbaliser, which translates the geometric data into natural language, a pipeline we have explored empirically in the context of personalised outdoor route descriptions \cite{ilyankou_geospatial_2025}. The LLM Verbaliser is constrained by the Policy Enforcement module that enforces template-bounded outputs. This ensures that mandatory disclosures (e.g., `This route is sponsored') cannot be omitted, and the LLM can only paraphrase within strict safety limits.

Then, a Telemetry \& Provenance module generates structured metadata for every route segment, recording signals such as `data quality', `safety risk', or `commercial incentive'. This data flows into an Audit Log, which stores the generated route alternatives, the selected weightings, and the final disclosures presented to the user. Such logging mechanism is essential for detecting EPs post-hoc, and allowing users, developers, or regulators to review discrepancies between the system's logic and its verbal explanations.

The final technology layer of compute and storage handles real-time symbolic pathfinding. The entire workflow adheres to strict standards for Auditability and Versioning (Figure \ref{fig:reference-implementation} sidebar). This ensures that as the routing algorithm evolves, its decision-making remains transparent and traceable for the end user.

\subsection{Operationalising the DP/EP Distinction via Route Comparison}

A key challenge in distinguishing DPs from EPs is that both can produce similar outcomes (i.e., a suboptimal route), but differ in origin. We propose that the Symbolic Routing Engine addresses this by computing not just the selected route but a set of Pareto-efficient baseline alternatives optimised along objective axes of time, distance, and elevation (in other words, compare directly with  `traditional' pathfinding). Such baselines can serve as a counterfactual reference against which the proposed route can be audited.

For each suggested route, the system computes a \emph{detour cost} relative to the fastest baseline and an estimated third-party benefit (e.g., footfall value derived from the `commercial incentive' flag). Where detour cost is non-trivial and third-party benefit is positive, the asymmetry is a structured signal of potential hidden steering, which is a DP by routing origin, and such signal can deterministically trigger the sponsorship disclosure. Conversely, where route quality degrades due to sparse or low-confidence data (flagged via `uncertainty score'), without any corresponding third-party benefit, the system classifies the failure as an EP and triggers uncertainty-aware hedging instead.

Therefore, not all routes through partner zones constitute deception. Where the proposed route incurs no detour cost relative to the baseline, the commercial flag does not imply user harm. The disclosure policy can be tiered: when detour cost is negligible, a lightweight ambient acknowledgement suffices (`This area includes partner businesses'); only where detour cost is material does the system trigger full sponsorship disclosure such as `This route adds 6 minutes and passes through a Partner Zone'. This graduated approach preserves warning salience by reserving assertive disclosure for cases where asymmetry between user cost and third-party gain is detectable.

This cost/benefit asymmetry serves as a computable proxy for intent: deception tends to produce routes where user cost and vendor gain co-occur, while unintentional error produces user cost without beneficiary. By being explicit about such a trade-off, the system makes the asymmetry legible to the user, to auditors, and to regulators.

\subsection{Limitations}

First, subjective routing metrics such as `safety', `attractiveness' or `neighbourhood wealth' require proxy data such as crime statistics, street-level imagery judgements \cite{malekzadeh_urban_2025,wang_can_2025}, or retail composition \cite{ilyankou_supermarket_2023}, which all carry their own biases. For example, a safety score derived from crime data may encode racial or socioeconomic disparities. Beyond bias, these proxies face a validity problem: a high walkability score does not guarantee a user will experience a route as attractive, meaning the symbolic layer may operate on metrics that are verifiable in themselves but imperfectly representative of the subjective qualities they proxy. The architecture makes such discrepancies auditable, but does not resolve them.

Second, errors introduced by the LLM Interpreter, such as mistaking `quiet' as low-traffic rather than low-crime, or confusing `scenic' with `green' when the user meant `historic' propagate into the Symbolic Routing Engine before Policy Enforcement has any opportunity to intervene. Unlike routing errors, which leave an auditable trace in the detour-cost calculation, spatial and semantic misinterpretation (LLMs carry substantial but unevenly distributed geospatial knowledge \cite{manvi_large_2024,ilyankou_quantifying_2024}) may produce a route that is objectively optimal for the wrong objective, with no anomaly signal to trigger disclosure. This makes interpreter errors structurally harder to detect than either DPs or EPs as currently defined, and suggests that intent clarification (i.e., asking the user to confirm the system's interpretation of ambiguous qualifiers before routing) may be a necessary additional intervention.

Third, the audit log assumes all relevant signals can be captured as structured metadata, but emergent harms in novel, out-of-distribution contexts (e.g., a `scenic' route during flooding) may not map to any pre-defined flags.

Fourth, the tiered disclosure policy requires empirical calibration, likely on a personal level: what constitutes a negligible detour depends on the user's physical fitness and mobility; similarly, tolerance to repeated warnings before they become 'white noise' likely varies by individual.

Fifth, the Policy Enforcement module is configured by the operator deploying the system; a bad-faith operator could simply suppress mandatory disclosures, meaning the architecture's honesty guarantees are only as strong as the institutional constraints surrounding it.

However, all these are directions for future work. Our core claim that a neuro-symbolic architecture is more auditable than end-to-end LLM navigation holds even under these caveats.

\section{Conclusion}

The future of navigation is likely conversational, but conversation is an art of persuasion. As we transition from tool-like maps and navigators to agent-like guides, we risk walking into a `mirage' where commercial interests and hallucinations are hidden behind friendly dialogues. To prevent this, conversational navigation must adopt seamful principles that expose incentives and uncertainty rather than conceal them. Our proposed $2\times2$ framework clarifies where harms originate, whether deliberate or accidental, and informs interventions such as sponsorship disclosure and trust calibration via uncertainty-aware language. Combined with a neuro-symbolic architecture that grounds routing in verifiable logic -- where detour cost co-occurring with third-party gain deterministically triggers disclosure, making the DP/EP distinction computable -- these measures operationalise trust in navigation systems that admit, clearly and conversationally, when they are guessing and why.

\section{Acknowledgments}

This work was supported by Ordnance Survey \& UKRI Engineering
and Physical Sciences Research Council [grant no. EP/Y528651/1].

\bibliographystyle{ACM-Reference-Format}
\bibliography{references}

\end{document}